\documentclass[11pt,twoside,onecolumn]{article}
\newcommand{\be}{\begin{equation}}
\newcommand{\ee}{\end{equation}}
\def\ba{\begin{eqnarray}}
\def\ea{\end{eqnarray}}

\def\o{\omega}
\def\f{\frac}
\def\({\left(}
\def\){\right)}
\def\ls{\left[}
\def\rs{\right]}
\def\lc{\left\{}
\def\rc{\right\}}

\title{\bf Vacuum energy momentum tensor in (2+1) NC scalar field theory}
\author{P. Nicolini\thanks{nicolini@cmfd.univ.trieste.it}$\ $
\\
\\ {\em Dipartimento di Scienze Matematiche dell'Universit\`a degli
Studi di Trieste and}\\{\em  INFN, Sezione di Trieste, Via Valerio
12, 34127 Trieste, Italy}}

\begin{document}
\maketitle


\begin{abstract}

A scalar field in $(2+1)$ dimensional Minkowski space-time is
considered. Postulating noncommutative spatial coordinates, one is
able to determine the (UV finite) vacuum expectation value of the
quantum field energy momentum tensor. Calculation for the $(3+1)$
case has been performed considering only two noncommutative
coordinates. The results lead to a vacuum energy with a lowered
degree of divergence, with respect to that of ordinary commutative
theory.
\end{abstract}

\raggedbottom
\setcounter{page}{1}
%
%
\setcounter{equation}{0}
%
%

%

%
Recent cosmological data \cite{cosmo} show that the expansion of
the universe is presently accelerating.  This acceleration effect
is driven by a still undetermined sort of dark mass/energy. It is
a common belief that the cosmological constant, or vacuum energy,
provides a  relevant contribution to the dark energy \cite{linde}.
On the contrary of what was believed in the past, the cosmological
constant is not zero but gives a contribution to the total mass
energy of the universe, which is comparable to the one  from
visible matter. By means of the value of $\Lambda$, coming from
observations of Type Ia supernovae, one can estimate the
cosmological energy density to be \cite{kir} \be \left|{\cal
E}_\Lambda \right| =\left|\Lambda\right|/(8\pi G)\sim 5\times
10^{-31}g/cm^3. \label{coscon}\ee Since the Zel'dovich seminal
paper \cite{zel}, the cosmological constant is interpreted as the
macroscopic effect produced by quantum matter vacuum fluctuations.
Unfortunately, quantum field theory and even string theory are
presently unable to reproduce the aforementioned value
(\ref{coscon}). Technically, the main difficulty encountered is
related to the bad short distance behavior of point-like objects,
which is only partially cured in string theory. On the other hand
the great interest towards noncommutative geometry \cite{ncg}
could be explained with the hope that a correct formulation of
noncommutative field theory could be free from UV divergences. In
this framework one expects to find a finite value for the vacuum
energy and thus for the cosmological constant. Many authors faced
the problem of obtaining the vacuum expectation value for the
energy momentum tensor (EMT) in a noncommutative space-time
\cite{ncemt}: at present for (3+1) dimensional fields there are
theories based on the Moyal $\star$-product, which are notoriously
affected by some important problems, such as unitarity, Lorentz
invariance breaking and UV/IR mixing. Exploiting the coherent
states formalism,  an alternative approach has been proposed
\cite{anais} to formulate a (2+1) dimensional noncommutative field
theory, which exhibits some positive aspects: the formalism is
handy, the UV divergences can be cured, a minimal length is
introduced in a clear and evident way. At the light of the above
considerations, we retain useful to face the problem of the vacuum
energy due to a scalar field in (2+1) dimensional space-time, when
only spatial coordinates noncommute, while the time is still
commutative. This is a preliminary step to a more realistic
computation in (3+1) dimensions.

The expression for the EMT is generally given by \be
T_{\alpha\beta}=\phi_{,\alpha}\phi_{,\beta}-\f{1}{2}\eta_{\alpha\beta}
\eta^{\lambda\delta}\phi_{,\lambda}\phi_{,\delta}+\f{1}{2}m^2\phi^2
\eta_{\alpha\beta}. \label{emt} \ee After expansion on Fourier
modes, one can compute its quantum expectation value on Fock
space. Special interest attaches to the vacuum state $|0\rangle$,
for which one can write \cite{bd} \be \langle
0|T_{\alpha\beta}|0\rangle =\int d{\bf k}\ \
T_{\alpha\beta}[u({\bf k}), u^\ast({\bf k})] \ee where
$T_{\alpha\beta}[u({\bf k}), u^\ast({\bf k})]$ denotes the
bilinear expression (\ref{emt}) for $T_{\alpha\beta}$ and \be
u({\bf k})=\f{1}{\sqrt{2\(2\pi\)^{2} k^0}}\exp\(-ik_\mu x^\mu\),
\ee with the wave vector $k^\mu =(k^0, {\bf k})$. The temporal
components, namely $\langle T_{00}\rangle $, gives the vacuum
energy density \be {\cal E} =\int d{\bf k}\ \ ({\bf k}^2 +m^2)\
u({\bf k})u^\ast({\bf k}) \ee a quantity that is non surprisingly
different from zero. It is well known that to the zeroth order in
the coupling, the field theory corresponds to an infinite
collection of harmonic oscillators with the $n^{th}$ oscillator
having a zero-point, or ground state energy, given by
$\f{1}{2}\o_n$ and the sum $\sum_n\f{1}{2}\o_n$ diverges. This
represents a well known and open problem when gravity is present,
because the standard procedure of (infinite) rescaling of ground
state energy is no more viable. Regarding this problem
noncommutative geometry could provide a natural UV cut-off, namely
a minimal length in space-time.

The simplest NC manifold is a 2D plane with spatial coordinate
operators $\hat{x}_i$ subject to the following relations \be \ls
\hat{x}_i, \hat{x}_j\rs =
i\theta\epsilon_{ij}\,\,\,\,\,\,\,i,j=1,2\label{ncrules}\ee
$\theta$ has dimension of a length squared and measures the
noncommutativity of spatial coordinates, while the time like
coordinate $x_0$ remains commutative.  As a consequence of
(\ref{ncrules}) the noncommutative plane is divided into plaquette
of area $\theta$. One cannot speak of points anymore and the space
becomes blurry. Then the best we can do is to define average
coordinates $x_i$ in an appropriate coherent state basis. Assuming
that momenta commute among themselves, the Fourier modes
  exhibit an additional term responsible of a gaussian
dampening factor \cite{anais} \be u({\bf
k})=\f{1}{2\pi}\f{e^{-ik_\mu x^\mu -\theta{\bf
k}^2/4}}{\sqrt{2k^0}}. \label{modes} \ee Now we are ready to
define a quantum field on a noncommutative plane in terms of the
above modes \be \phi(x)=\f{1}{2\pi}\int \f{d{\bf k} e^{-\theta
{\bf k}^2 /4}}{\sqrt{2k^0}}\ls e^{ik_\mu x^\mu}a^{\dagger}({\bf k}
) + e^{-ik_\mu x^\mu}a({\bf k})  \rs \label{ncfour} \ee where
$a^{\dagger}({\bf k})$, $a({\bf k})$ are usual
creation/annihilation operators on Fock states with definite
energy and momentum \cite{anais}.  Substituting the explicit form
for $u({\bf k})$ given in (\ref{modes}) one is left with the
following integral \be {\cal E}_\theta =\f{1}{8\pi^2}\int d{\bf k}
\ \ e^{-\theta {\bf k}^2 /2}\ \ \sqrt{{\bf k}^2+m^2}. \ee The
calculation leads to the following result \be {\cal E}_\theta
=\f{1}{8\pi\theta^{3/2}}\lc \theta^{1/2} m + e^{\theta
m^2}\f{\sqrt{\pi}}{2} Erfc\( \theta^{1/2} m\)\rc \ee where
$Erfc(x)$ is the  complementary error function \cite{grad}. For
the massless field one obtains \be {\cal E}_\theta=
\f{1}{8\pi\theta^{3/2}}\f{\sqrt{\pi}}{2}.\ee
 We stress that ${\cal E}_\theta$ is finite and
does not need any renormalization procedure.


The above calculation can be repeated in the $(3+1)$ dimensional
case. Of course one is naturally tempted to consider a full
noncommutative space-time in which each pair of coordinates
mutually anticommutes according to an homogeneous minimal length
set by an unique $\theta$ parameter. A natural prescription could
be given by a generalization of (\ref{ncrules}) to a
multidimensional (anti)commutator in a similar fashion to what has
been argued by Nambu in one of his famous papers \cite{nambu} \be
\ls \hat{x}_\alpha, \hat{x}_\beta, \hat{x}_\gamma, \hat{x}_\delta
\rs = \theta \epsilon_{\alpha\beta \gamma\delta}. \ee In absence
of a viable field theory based on commutation relation of this
kind one is left with the ``binary'' noncommutativity coming from
rule (\ref{ncrules}), which forces the space-time to be foliated
among noncommutative planes. Unwillingly we follow the usual
wisdom to keep time as a commutative coordinate in order to avoid
problems with unitarity \cite{unit}. So the best we can consider
is still the noncommutativity in a plane, for instance the $(x_1,
x_2)$ plane. It is of course unpleasant having to do with a
inhomogeneous space-time, even if such assumption can be accepted
because the loss of homogeneity regards only the extremely high
energies in which noncommutativity should play his fundamental
role. Furthermore the Lorentz invariance breaking is an occurrence
that cannot be excluded for so small scales of length. Mimicking
calculations of the (2+1) dimensional case one adopts the
following modes \be u({\bf k})=\f{1}{\sqrt{2\(2\pi\)^{3}
k^0}}e^{-ik_\mu x^\mu-\theta\(k_1^2 +k_2^2\) } \ee to find \be
{\cal E}_\theta =\f{1}{16\pi^3}\int \ dk_3 \ d{\bf q} \
\sqrt{k_3^2 +{\vec{q}}^2 +m^2} \exp(-\theta{\vec{q}}^2)
\label{tre} \ee where ${\bf k}=(k_1, k_2, k_3)$ and ${\bf
q}\equiv(k_1, k_2)$. Then one obtains  \be {\cal E}_\theta
=\f{1}{16\pi^2\theta^{3/2}}\int_{-\infty}^\infty \ dk_3
\lc\theta^{1/2}\(k_3^2 +m^2\)^{1/2} +\f{\sqrt{\pi}}{2} \
e^{\theta\(k_3^2 +m^2\)} \  Erfc\(\theta^{1/2}(k_3^2
+m^2)^{1/2}\)\rc . \ee For the massless field the analogous
expression is  \be {\cal E}_\theta
=\f{1}{16\pi^2\theta^{3/2}}\int_{-\infty}^\infty \ dk_3
\lc\theta^{1/2}k_3 +\f{\sqrt{\pi}}{2} \ e^{\theta k_3^2} \
Erfc\(\theta^{1/2} k_3\)\rc . \ee Even if the energy results to be
clearly not finite, we can observe that the degree of divergence
has been reduced to the one of a (1+1) dimensional theory. This is
evident looking at the large argument behavior of the integrand,
in which the Erfc term is suppressed \cite{grad} \be
\f{\sqrt{\pi}}{2} \ e^{\theta x^2} \ Erfc\(\theta^{1/2} x\)\sim
\f{1}{2\sqrt{\pi}}\sum_{k=0}^{n-1}\f{(-1)^{k}\Gamma(k+1/2)}{(\theta
x^2)^{k+1/2}}+O\(\f{1}{(\theta x^2)^{n+1/2}}\) \ \ \ \ \ large \ \
\ x \ee where $x$ is equal to $k_3$ or $(k_3^2 +m^2)^{1/2}$
depending on the case. Then the asymptotic value of the energy
density is \be {\cal E}_\theta
\sim\f{1}{8\pi^2\theta^{3/2}}\int^\infty \ dk_3 \ \(
\theta^{1/2}\sqrt{k_3^2 +m^2}+\f{1}{2 \ \theta^{1/2}\sqrt{k_3^2
+m^2}}\) \ee and analogously for the massless case.  We have got
the divergence relative to the remaining (commutative) coordinates
$x_3$. Such divergence can be cured through customary
renormalization methods.

In this letter we adopted a model, that, even if simplified, is
able to give a finite vacuum energy for a (2+1) scalar field and
lowers the degree of divergence of the (3+1) theory down to the
degree of divergence of an effective (1+1) field theory. It would
be interesting to deal with a nonrenormalizable theory since the
beginning, rather than with the ``good'' scalar field theory. The
hope would be to obtain (1+1) theory, whose divergences can be
removed via usual renormalization procedures. Finally, we are left
with the need of extending the present (2+1) method to higher
dimensions in order to perform a more realistic calculation of the
vacuum energy density.

\subsection*{Acknowledgements} The author is grateful to Euro
Spallucci for fruitful discussions and to Thomas Schucker for the
hospitality at CPT, Marseille during a stay supported by CNR, the
Italian national research council.


\begin{thebibliography}{99}
%
\bibitem{cosmo}R.A. Knop et al., astro-ph/0305008; J.L. Torny et al.,
Astrophys. J. 594, 1 (2003); S. Perlmutter et al., Astrophys. J.
517, 565 (1999); A.G. Riess et al., Astron. J. 116, 1009 (1998);
J. L. Sievers et al., astro-ph/0205387; J. R. Bond et al.,
astro-ph/0210007; D. N. Spergel et al., astrp-ph/0302209.

\bibitem{linde} J. Kratochvil, A. Linde, E. V. Linder and M.
Shmakova, astro-ph/0312183.

\bibitem{kir} R. P. Kirshner, Proc. Natl. Acad. Sci. USA, 96, 4224
(1999).

\bibitem{zel} Ya. B. Zel'dovich,
Zh. Eksp. \& Teor. Fiz. Pis'ma 6, 883 (1967). English translation
in Sov. Phys. - JEPT Lett. 6, 316.

\bibitem{ncg}A. Connes, Comm. Math. Phys. 155, 109 (1996);
N.~Seiberg and E.~Witten , JHEP \  9909, 032 (1999);
 M. Douglas, N. A. Nekrasov, Rev. Mod. Phys. 73, 977
(2001); R. J. Szabo, Phys. Rept. 378, 207 (2003);  Chong-Sun Chu,
J. Lukierski, W. J. Zakrzewski, Nucl.Phys. B632, 219 (2002).



\bibitem{ncemt}  A. Gerhold,
J. Grimstrup, H. Grosse, L. Popp, M. Schweda, R. Wulkenhaar,
hep-th/0012112;  Y. Okawa, H. Ooguri, hep-th/0103124; M.
Abou-Zeid, H. Dorn, Phys.Lett. B514 183, (2001);  J. M. Grimstrup,
B. Kloiböck, L. Popp, V. Putz, M. Schweda, M. Wickenhauser,
hep-th/0210288; A. Das, J. Frenkel, Phys.Rev. D67, 067701 (2003);
S.  Bellucci, I. L. Buchbinder, V. A. Krykhtin, Nucl. Phys. B665,
402 (2003).

\bibitem{anais}
A. Smailagic and E. Spallucci,
J.Phys. A36,  517, (2003).

\bibitem{bd}
N.D. Birrell \& P.C.W. Davies, {\em Quantum fields in curved space},
Cambridge, Uk: Univ. Pr. (1982).


\bibitem{nambu}
Y. Nambu,
Phys. Rev. D7, 2405 (1973).

\bibitem{unit} M. Chaichian, A. Demichev, P. Presnajder, A. Tureanu,
Eur. Phys. J. C20, 767 (2001).


\bibitem{grad}
I.S. Gradshteyn and I.M. Ryzhik, {\em Table of Integrals, Series
and Products}, Academic Press, (1965); M. Abramowitz  \& I. A.
Stegun, {\em Handbook of mathematical functions}, Dover
Publication, Inc., New York (1972).


%
\end{thebibliography}
\end{document}